\documentclass[reprint,amsmath,amssymb,aps,prb,superscriptaddress]{revtex4-2}

\usepackage{graphicx}
\usepackage{bm}
\usepackage[colorlinks=true, linkcolor=blue, citecolor=blue, urlcolor=blue]{hyperref}
\usepackage{siunitx}
\usepackage{xcolor}

\newcommand {\ii}          {\mathrm{i}}
\newcommand {\ket}[1]      {\lvert#1\rangle}
\newcommand {\isopz}       {\tilde{\mathcal{P}}_z}
\newcommand {\idmat}       {\mathbf{1}}   
\providecommand {\micro} {\textmu}
\newcommand{\Vbg}{V_{\mathrm{bg}}}
\newcommand{\Vtg}{V_{\mathrm{tg}}}
\newcommand{\nP}{n^\mathrm{P}}
\newcommand{\nQ}{n^\mathrm{Q}}
\newcommand{\Rxx}{R_{xx}}
\newcommand{\Rxy}{R_{xy}}
\newcommand {\kdotp}   {\ensuremath{\mathbf{k}\cdot\mathbf{p}}}

\begin{document}

\title{Symmetry induced spin splitting of topological surface states at equal carrier density}
\author{Wouter Beugeling}
\email{wouter.beugeling@uni-wuerzburg.de}
\author{Shyam Sundar Yadav}
\author{Maximilian Hofer}
\author{Valentin L. Müller}
\author{Moritz Siebert}
\author{Tobias Kießling}
\affiliation{Experimentelle Physik III, Physikalisches Institut, Universität Würzburg, Am Hubland, 97074 Würzburg, Germany}
\affiliation{Institute for Topological Insulators, Universität Würzburg, Am Hubland, 97074 Würzburg, Germany}
\author{Ewelina M. Hankiewicz}
\affiliation{Institut für Theoretische Physik und Astrophysik, Universität Würzburg, Am Hubland, 97074 Würzburg, Germany}
\author{Hartmut Buhmann}
\author{Laurens W. Molenkamp}
\email{laurens.molenkamp@uni-wuerzburg.de}
\affiliation{Experimentelle Physik III, Physikalisches Institut, Universität Würzburg, Am Hubland, 97074 Würzburg, Germany}
\affiliation{Institute for Topological Insulators, Universität Würzburg, Am Hubland, 97074 Würzburg, Germany}

\date{November 13, 2026}

\begin{abstract}
Surface states are among the most striking consequences of band inversion in topological materials.
In magnetotransport of three-dimensional topological insulators, the typical signature of surface state transport is a quantized Hall conductance with nontrivial sequences of integer filling factors, varying with the total carrier density.
For two independent Dirac carriers, one expects Hall plateaus only at odd-integer filling factors.
In this work, we report the observation of symmetry induced spin splitting in magnetotransport experiments on a dual-gate three-dimensional topological insulator device based on HgTe.
In the topological surface state regime, we observe Hall plateaus at both odd and even integers when we gate the device symmetrically, i.e., with equal density at the top and bottom.
This phenomenon originates in structural inversion symmetry, which remarkably induces a spin splitting and forces the eigenstates to be symmetric and antisymmetric linear combinations of the top and bottom surface states. This is in contrast with two-dimensional systems, where spin splitting is caused by absence of inversion symmetry.
We interpret the experimental observations with a perturbative model using isoparity eigenstates, the conserved quantum number associated with structural inversion.
The result demonstrates that the picture where the top and bottom states are treated as independent Dirac carriers is not valid when the carrier density is symmetric.
\end{abstract}

\maketitle

\section{Introduction}
\label{sec:introduction}

Surface-state transport is a key feature of three-dimensional topological insulators (3D TIs) \cite{Konig,Fu3DTI,HasanRMP,Xu2014}.
As originally predicted in 1985 \cite{VPS, KusmartsevTsvelik1985}, the band inversion occuring at the interface between a band inverted and normal material leads to two-dimensional Dirac-type surface states at the interface.
Magnetotransport experiments of 3D TI devices have confirmed the existence of these surface states through observation of the quantum Hall effect. These experiments typically exhibit clear signatures of multi-carrier behavior, such as the absence of certain Hall plateaus \cite{BruneQHE, Brunescreening, Xu2014}, caused by the presence of at least two types of carriers, namely the surface states at top and bottom interface.
In some configurations, the presence of further carrier types, such as massive Volkov-Pankratov states, can be observed \cite{MahlerVPS,Lixian}.

The electrostatics of these devices is an essential aspect of the physics, but is relatively intricate, due to screening effects of the carriers in the system \cite{Brunescreening,MahlerVPS,HoferEASelfcon}: When for example the device is equipped with a top gate, the gate voltage affects the top surface more strongly than the bottom surface.
Thus, gating generally affects the carrier population asymmetrically, breaking structural inversion symmetry. For two-dimensional devices, structural inversion asymmetry generally manifests itself as Rashba spin-orbit coupling, and this also holds for two-dimensional topological insulator (2D TI) materials \cite{ZhangEA2001,GuiEA2001,GuiEA2004PRB,GuiEA2004EPL,HinzEA2006}.

In contrast, for 3D TI devices, the dominant effect is an energy splitting between the two surface states, which both individually are not spin split because they are nondegenerate.
The relevant physics for symmetric gating in 3D TIs, which requires more experimental effort than asymmetric gating (a dual-gate heterostructure where the top and bottom gate voltages can be tuned independently), has not been fully explored so far. Naively, one may try to understand this case in a basis of independent top and bottom surface states, where both surface states are described as independent Dirac fermions. This should then result into an odd integer sequence of Hall plateaus at equal density. Our key finding is that this picture is not valid: In a symmetrically gated 3D TI system in a finite perpendicular magnetic field, the actual eigenstates are formed by symmetric and antisymmetric superpositions of the top and bottom surface states. Hall plateaus occur at all integers, not only the odd ones.

The origin of this effect is the presence of structural inversion symmetry.
If bulk-inversion asymmetry is negligible, which usually is a good approximation in the context of magnetotransport \cite{Weiler1981bookchapter,Winkler2003_book}, this symmetry leads to a conserved quantum number, which we name isoparity \cite{Isoparity,FundamentalSymmetries}.
The isoparity operator acts as parity in the spatial degrees of freedom, i.e., by reflection in the $z$ direction, and acts nontrivially in the orbital degrees of freedom, determined by the parity of the atomic orbitals ($s,p_x,p_y,p_z$) and the spin.
This operator defines a good quantum number even at finite momenta, unlike the orbital or spin angular momentum operators  $J_z$ and $S_z$ \cite{Isoparity}.

In this work, we report the observation of the consequences of structural inversion symmetry in a topological insulator device. We have performed magnetotransport experiments on a $70\,\mathrm{nm}$ thick layer of HgTe, a three-dimensional topological insulator, equipped with gates on top and bottom, and fabricated into a Hall bar geometry.
By tuning both gate voltages independently, at a constant finite magnetic field, we map out the longitudinal and transversal resistance of the device.
The result is an intricate pattern of quantized Hall plateaus and transitions, as we show in Fig.~\ref{fig:dualgate}.
If we take a cut through the data at structural symmetry (i.e., where the carrier density profile is symmetric) in the regime where surface-state transport dominates, the Hall conductance goes through a sequence of plateaus at integer multiples of $e^2/h$, contradicting the naive picture of two independent Dirac fermions at the surfaces (cf.\ Ref.~\cite{Xu2014}).

In order to theoretically interpret these features, we perform in-depth analysis of the Landau levels by means of $\mathbf{k}\cdot\mathbf{p}$ theory \cite{Kanemodel,Novik,kdotpy}.
Based on these results, we propose a perturbative effective model involving one pair of Landau levels, in a basis of isoparity $\pm 1$ eigenstates, acting as two pseudospin states.
(Total angular momentum $J_z$ and proper spin $S_z$ are not suitable for this purpose, because they are not conserved.)
The model involves two perturbative terms that lift the degeneracy, one symmetric and one antisymmetric in isoparity, which model the magnetic field \cite{Isoparity} and electric field \cite{FundamentalSymmetries}, respectively.
The perturbative model fits accurately to the experimental conductance map, establishing the following physical picture:
The magnetic field induces a Zeeman-like splitting between the two states with the same Landau index (see Fig. \ref{fig:ll}(a)).
For symmetric gating (i.e., a symmetric charge carrier distribution) the splitting is purely Zeeman-like and there is no Rashba effect. Only a sufficiently strong asymmetry in the electrostatic potential can overcome the Zeeman splitting and force the eigenstates to the top and bottom surfaces.
Our perturbative model with competing terms explains both limits: On the one hand, the splitting in the symmetric case (i.e., for equal carrier density at top and bottom) leads to the even-and-odd filling-factor Hall plateaus.
On the other hand, no splitting occurs between the plateau transitions in the asymmetric-density regime, because states with unequal Landau indices do not mix. The observed spin splitting at equal density is thus the result of a protected symmetry, i.e., structural inversion asymmetry, with isoparity as its conserved quantum number.

We emphasize that this mechanism is indeed very different from the physical mechanisms that give rise to parity jumps in quantum oscillations and Rashba splitting of two-dimensional systems: In two-dimensional systems, the field induced by the gate potential acts predominantly by the Rashba effect, leading to different densities in the spin-orbit split states which can be observed as beating patterns in Shubnikov-de Haas oscillations and missing plateaus in the Hall resistance \cite{ZhangEA2001,GuiEA2001,GuiEA2004PRB,GuiEA2004EPL,HinzEA2006}.

\section{Results}
\subsection{Magnetotransport of the dual-gate device}
\label{sec:magnetotransport}

\begin{figure*}[t]
\includegraphics[width=154mm]{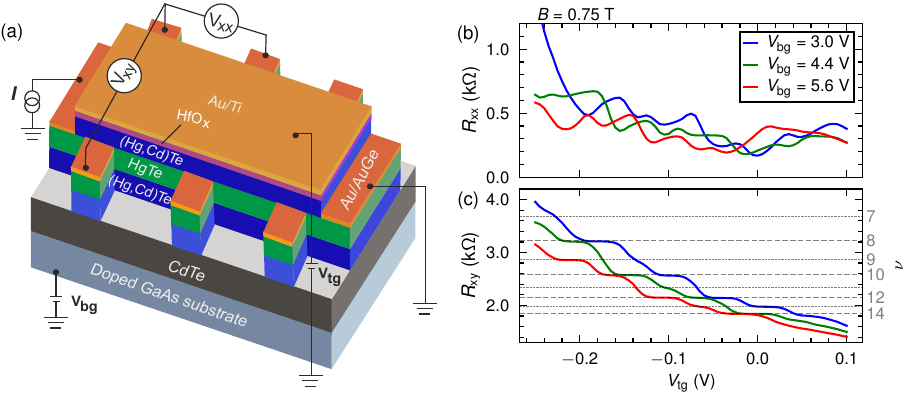}
\caption{\label{fig:sampleraw}%
Magnetotransport measurement in a HgTe heterostructure.
(a) Schematic of the device structure and measurement configuration (dimensions not to scale). The device is a dual gated Hall bar, with a $70\,\mathrm{nm}$ thick HgTe layer as the topological insulator. The GaAs substrate acts as the bottom gate and the Ti/Au layer as the top gate. The HfO$_{x}$ insulating dielectric separates the topological insulator structure from the top gate.
(b) Top gate dependence of longitudinal resistance $\Rxx$ and
(c) Hall resistance $\Rxy$ for several bottom gate voltages (shown for $\Vbg = 3.0, 4.4$, and $5.6\,\mathrm{V}$) under constant perpendicular magnetic field $B=0.75$\,T and temperature $T=150$\,mK. The total filling factors $\nu$ are determined from $\Rxy = h/\nu e^2$.}
\end{figure*}

We have performed magnetotransport experiments on a 70 nm thick HgTe layer grown by MBE on a virtual CdTe substrate \cite{LeubnerEA2016} on top of a Si-doped GaAs wafer for bottom gate applications.
We have used an eight-terminal Hall bar design with length between voltage probes $\SI{540}{\micro\meter}$ and width  $\SI{180}{\micro\meter}$ for the magnetotransport measurements. For the top gate, we have used a $15\,\mathrm{nm}$ thick $\mathrm{HfO_x}$ dielectric and a $10\,\mathrm{nm}/100\,\mathrm{nm}$ Ti/Au gate electrode, see Fig.~\ref{fig:sampleraw}(a) and SI Appendix, section \ref{sec:supp_device} and Fig.~\ref{figsupp:deviceimage}.
The magnetotransport measurements have been carried out using a low-frequency AC lock-in technique (see SI Appendix, section~\ref{sec:supp_transport} for details), with simultaneous measurements of the longitudinal ($\Rxx$) and transverse ($\Rxy$) resistances. The excitation current has been measured from the voltage drop across a calibrated series resistor. The measurements are performed in a $^{3}$He-$^{4}$He dilution refrigerator at a nominal base temperature of $150\,\mathrm{mK}$.

In Figs.~\ref{fig:sampleraw}(b) and (c) we show the top gate ($\Vtg$) dependence of the measured $\Rxx$ and $\Rxy$ for different values of bottom gate voltage $\Vbg= 3.0, 4.4$, and $5.6\,\mathrm{V}$, under a constant perpendicular magnetic field of $B=0.75\,\mathrm{T}$. (Further data in SI Appendix, Fig.~\ref{figsupp:RxxRxy}.) Owing to the high quality of the sample, Shubnikov-de Haas oscillations and Hall quantization are observed even at relatively low magnetic fields ($< 1\,\mathrm{T}$). The filling factor $\nu$ is obtained from the quantized Hall resistance
$\Rxy = h/\nu e^2$.
For $\Vbg = 3.0\,\mathrm{V}$, not all integer filling factors are observed, with $\nu = 8, 9, 10, 12, 13$ present. Similarly, the observed sequences are $\nu = 8, 10, 11, 12, 14$ at $\Vbg = 4.4\,\mathrm{V}$, and $\nu = 9, 10, 12, 13, 14$ at $\Vbg = 5.6\,\mathrm{V}$.
The absence of some filling factors is attributed to the interplay between the top and bottom surfaces and to screening effects as  previously observed in HgTe 3D TI devices \cite{BruneQHE,Brunescreening,MahlerVPS,HoferEASelfcon}.

\begin{figure*}[t]
\includegraphics[width=140mm]{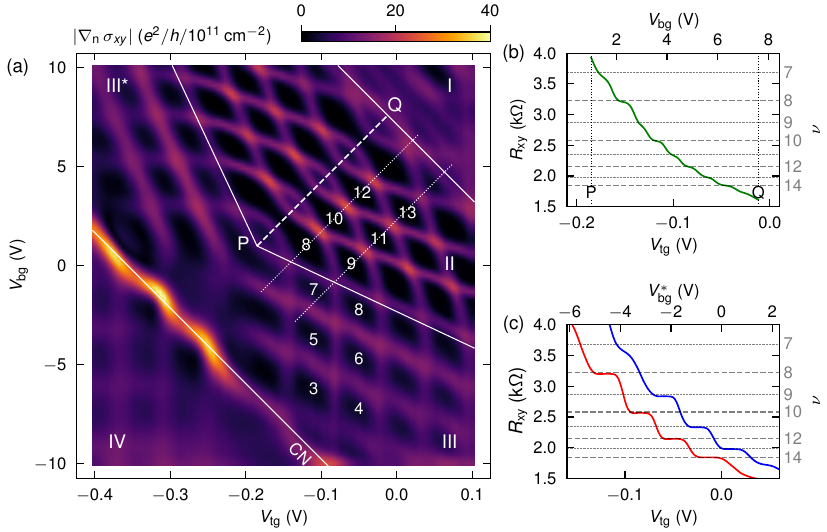}
\caption{\label{fig:dualgate}%
Dual-gate scan of the Hall conductance.
(a) The quantum Hall plateau structure as function of top and bottom gate voltage, $\Vtg$ and $\Vbg$. The axes have been scaled such that a horizontal and vertical distances correspond to the same change in total carrier density. The color scale represents the weighted gradient $|\nabla_n \sigma_{xy}|$, i.e., the magnitude of the derivative in the plotted coordinates. The dark areas, where this derivative is zero, correspond to the Hall plateaus. The integer values $\nu$ indicate the filling factors, where $R_{xy}=h/\nu e^2$.
Region II is the regime of surface-state dominated transport, whereas elsewhere parallel conduction of bulk electrons or holes occurs.
The plateau structure is symmetric with respect to the line P--Q, which corresponds to equal density at top and bottom.
The line labeled `CN' indicates overall charge neutrality and separates $p$- and $n$-type transport.
(b) Hall resistance $\Rxy$ as function of the gate voltages along the equal-density line P--Q.
A complete sequence of integer filling factors $\nu = 7, 8, 9, \ldots, 14$ is observed, where the plateaus at the odd integer filling factors ($\nu = 9, 11, 13$) are resolved more weakly than the even ones.
(c) Hall resistance $\Rxy$ along the two dotted lines parallel to P--Q. The filling-factor sequences are even-only ($\nu=8,10,12,14$; red) and odd-only ($\nu=9, 11, 13$; blue). The quantity on the secondary horizontal axis is the relative back gate voltage $\Vbg^*\equiv (A_\mathrm{tg}/A_\mathrm{bg})\Vtg$.}
\end{figure*}

Furthermore, we can tune the device into different transport regimes (see also, e.g., Refs.~\cite{ZieglerQHE,HoferFuchsTransport}) by independently varying the gate voltages $\Vtg$ and $\Vbg$, as shown in Fig.~\ref{fig:dualgate}(a) for $B=0.75\,\mathrm{T}$. (For $B=1\,\mathrm{T}$ and $B=2\,\mathrm{T}$, see SI Appendix, Fig.~\ref{figsupp:other_magn_fields}.)
In view of the vastly different gate actions of top and bottom gates, the axes have been scaled such that a given distance on the horizontal and vertical axis corresponds to the same change in carrier density (see SI Appendix, section~\ref{sec:supp_data_analysis}).  
We visualize the pattern of Hall plateaus in terms of the derivative of $\sigma_{xy}$, more precisely as the magnitude of the weighted gradient
\begin{equation}\label{eq:weighted_gradient}
\nabla_n \sigma_{xy} = \left( \frac{1}{A_\mathrm{tg}} \frac{\partial \sigma_{xy}}{\partial \Vtg},\  \frac{1}{A_\mathrm{bg}} \frac{\partial \sigma_{xy}}{\partial \Vbg} \right),
\end{equation}
where $A_\mathrm{tg}$ and $A_\mathrm{bg}$ are the average gate actions of both gates, 
with $A_\mathrm{tg}/A_\mathrm{bg} \sim 38$. The darker areas, where $|\nabla_n\sigma_{xy}|=0$, correspond to Hall plateaus and bright features indicate the transitions.
They form an intricate distorted grid-like pattern, with a remarkable symmetry with respect to the diagonal axis P--Q, which we will later identify as the line where the densities at the top and bottom side of the device are equal, while the bulk is unoccupied.

We distinguish four transport regimes, labelled I--IV in Fig.~\ref{fig:dualgate}(a) (cf.\ Ref.~\cite{HoferFuchsTransport}).
The regions I and IV are strongly $n$- and $p$-type, respectively. In these regimes, bulk states of the conductance and valence band, respectively, are populated and contribute to transport, which leads to poorly defined Hall quantization.
The transition between $n$- and $p$-type transport (i.e., $\sigma_{xy}$ being positive and negative, respectively) occurs between regimes III/III* and IV. The charge neutrality line [labeled as `CN' in Fig.~\ref{fig:dualgate}(a)] that separates these regimes is perpendicular to the equal-density line.
In region II, the Fermi level is in the bulk gap, where only the topological surface states are responsible for transport.
In region III at the lower right corner, the total density is $n$-type but there is a strong density imbalance between the top and bottom of the sample. As we will demonstrate below, one of the surfaces is in the topological surface state regime, while for the other the top of the valence band is populated, more specifically the massive Volkov-Pankratov states \cite{MahlerVPS}.
Region III* in the upper left corner is equivalent, with top and bottom side exchanged.

\subsection{Structural inversion symmetry in the topological surface state regime}
\label{sec:tss}

In the regime where only the topological surface state are occupied, region II in Fig.~\ref{fig:dualgate}(a), we observe two distinct features.
Away from the equal density line P--Q, the Hall plateaus form a characteristic pattern of lozenges.
The pattern at the equal-density line is markedly different: The transitions do not intersect, but form avoided crossings, leading to peculiar eye-shaped features.
The transitions in this regime are all single Landau levels, i.e., the filling factors of adjacent plateaus (a pair of lozenges sharing an edge) differ by one. The transitions in region II are sharper than those in the other regions, because of the high mobility of the topological surface states compared to that of the other carrier types \cite{MahlerVPS,HoferFuchsTransport}.

The filling factors at the equal density line are the sequential integers $\nu = 7, 8, \ldots, 14, 15$. 
The $\Rxy$ dependence along P--Q [see Fig.~\ref{fig:dualgate}(b)] exhibits a clear even-odd effect, where the odd filling-factor plateaus are resolved less clearly than the even ones, 
as a result of their different origins:
The odd integer filling factors are the ones expected in a naive model of two independent Dirac-type surface states. The even filling factors occur at the avoided crossings.
Along other lines parallel to P--Q, we can find even-only (e.g., $\nu = 8, 10, 12, 14$) and odd-only (e.g., $\nu = 9, 11, 13, 15$) sequences, see Fig.~\ref{fig:dualgate}(c).

\begin{figure*}[t]
\includegraphics[width=178mm]{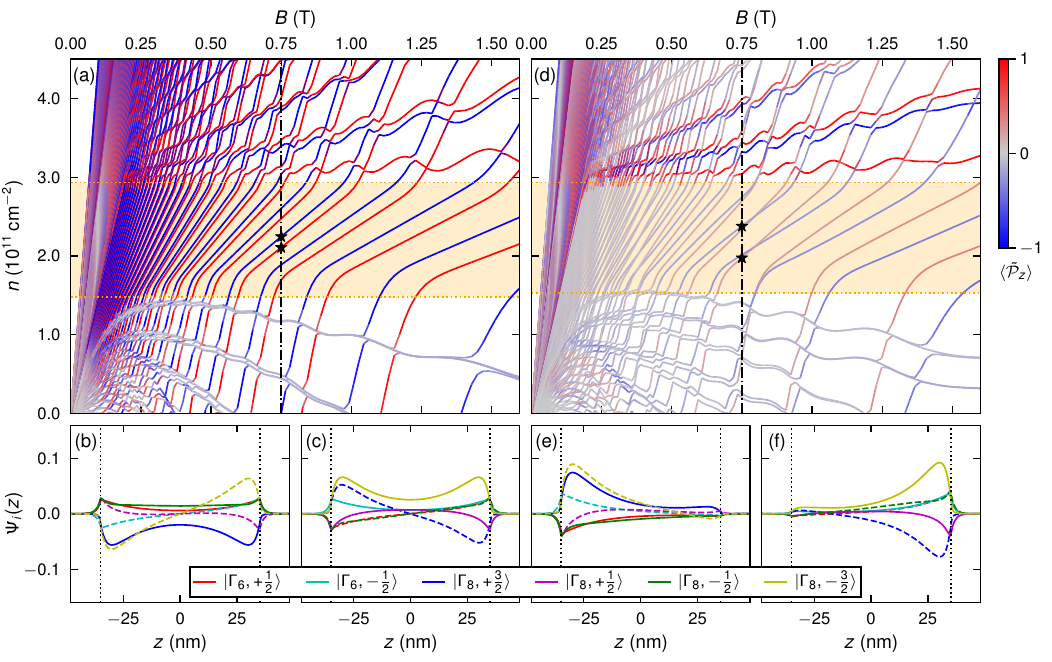}
\caption{\label{fig:ll}
Calculated Landau levels for 70 nm thick HgTe.
(a) The Landau levels (LLs) versus total carrier density, as a function of perpendicular magnetic field in the equal-density configuration.
The two stars mark a pair of LLs with identical LL index and with opposite isoparity expectation values $\langle\isopz\rangle= \pm1$.
The Hall filling factor in between these LLs is $\nu=12$.
The orange colored region highlights the topological surface-state regime.
(b), (c) Wave functions for the LLs marked in (a), expanded into individual orbital components $\ket{\Gamma_i,j_z}$ ($j_z=\pm\frac{1}{2},\pm\frac{3}{2}$) at a magnetic field of $B=0.75\,\mathrm{T}$. Solid curves indicate the real, dashed lines the imaginary parts. The small contributions from the $\ket{\Gamma_7,\pm\frac{1}{2}}$ orbitals are not shown.
(d) LLs in presence of an asymmetric electrostatic potential, modeled by a linear potential drop of $\Delta V = 5\,\mathrm{meV}$ across the HgTe layer.
The LLs marked by stars also enclose the $\nu=12$ plateau.
(e), (f) Wave functions for the LLs marked in (d).}
\end{figure*}

The crucial difference between the line P--Q and the remainder of region II is the presence of the inversion symmetry $z\to -z$ in the electron density.
For the $z\to-z$ symmetry of the Hamiltonian, with contributions of both crystallographic and electrostatic origins, isoparity is the relevant conserved quantity.
The isoparity operator is derived from representation theory of the point group that encodes the relevant crystal symmetry.
For unstrained bulk zincblende crystals, the relevant point group is $T_d$.
The Kane Hamiltonian $H_\mathrm{Kane}$ \cite{Kanemodel,Novik,kdotpy}, the established model for zincblende-type crystals, obeys the crystal symmetries by construction. 
The bulk inversion symmetry breaking terms in HgTe are insignificant, typically of the order of $\sim 1\,\mathrm{meV}$ \cite{Weiler1981bookchapter,Winkler2003_book}, and can be neglected in the modelling of many transport experiments.
Thus, we consider the inversion symmetric limit, which raises the symmetry to the point group $O_h$, which contains inversion $(x,y,z)\to(-x,-y,-z)$ as one of its elements, as well as the reflection $m_z: (x,y,z)\to(x,y,-z)$.

For heterostructures with growth direction (001), symmetric confinement and/or uniaxial strain reduces the point group from $O_h$ to $D_{4h}$.
In this geometry, we consider the reflection $m_z$ in particular.
It acts as the spatial parity transformation $P_z: z\to-z$ combined with a matrix multiplication by  \cite{Isoparity}
\begin{equation}
 R(m_z) = \mathop{\mathrm{diag}}(-\ii, +\ii, -\ii, +\ii, -\ii, +\ii, +\ii, -\ii)
\end{equation}
in the orbital degrees of freedom, expressed in the standard 8-orbital \kdotp{} basis, 
$\ket{\Gamma_6,+\frac{1}{2}}$, $\ket{\Gamma_6,-\frac{1}{2}}$,
$\ket{\Gamma_8,+\frac{3}{2}}$, $\ket{\Gamma_8,+\frac{1}{2}}$, 
$\ket{\Gamma_8,-\frac{1}{2}}$, $\ket{\Gamma_8,-\frac{3}{2}}$,
$\ket{\Gamma_7,+\frac{1}{2}}$, $\ket{\Gamma_7,-\frac{1}{2}}$.
For convenience, we define a real matrix operator
\begin{equation}\label{eq:q}
 Q = \ii R(m_z) = \mathop{\mathrm{diag}}(+1, -1, +1, -1, +1, -1, -1, +1)
\end{equation}
and define isoparity $\isopz=P_zQ$ as $P_z$ combined with multiplication by $Q$ in the orbital degrees of freedom \cite{Isoparity}.
The isoparity operator $\isopz$ thus commutes with the Kane Hamiltonian by definition. 

The full Hamiltonian $H=H_\mathrm{Kane} + V(z)$ also contains the potential term $V(z)$ from the electric field, induced both externally by the gates and internally from the carriers present in the system.
The transformation $m_z$ remains a symmetry of the full Hamiltonian if the potential is symmetric, i.e., $V(-z) = V(z)$.
If this equality is violated, the point group symmetry is reduced further to $C_{4v}$, because $m_z$ (together with all other transformations that do not preserve the $z$ coordinate) is no longer a symmetry of the full Hamiltonian.
We emphasize that inversion symmetry in a heterostructure device is a special case, that requires precise tuning of the population of the surface states by the two gates. For the electrostatic part $V(z)$, the symmetry is not protected by intrinsic properties of the system.

We confirm the role of isoparity by constructing a perturbative model, which involves tunable symmetry breaking, that we derive from analysis of Landau levels (LLs) in $\mathbf{k}\cdot\mathbf{p}$ band structure theory, using the aforementioned eight-orbital Kane model. 
We investigate the role of isoparity symmetry by comparing the LL spectra in two cases, one where this symmetry is conserved and one where it is broken by an electrostatic potential, see Figs.~\ref{fig:ll}(a) and (d), respectively.
The bulk gap with surface state transport only (region II) is visualized by orange color.
The symmetric case of Fig.~\ref{fig:ll}(a) corresponds to equal density at top and bottom surface, i.e., to the line P--Q of Fig.~\ref{fig:dualgate}(a).
The magnetic field in $z$ direction preserves the symmetry under reflection in $z$, so that isoparity $\isopz$ remains a good quantum number, but it lifts the degeneracy between $\isopz = \pm 1$ eigenstates \cite{Isoparity}.

The conservation of isoparity may be verified by inspection of the eigenstates shown in Figs.~\ref{fig:ll}(b) and (c).
In these figures, we show the decomposition of the eigenstates in terms of the envelope functions  $\psi_i(z)$ for each orbital $\ket{i}$ in the 8-orbital \kdotp{} basis (except $\ket{\Gamma_7,\pm\frac{1}{2}}$ which we have not explicitly shown).
The spatial component $P_z$ maps the envelope functions as $\psi_i(z) \to \psi_i(-z)$. Here, the envelope functions are either even or odd, $\psi_i(-z) = p_i\psi_i(z)$, with the parity eigenvalue $p_i=+1$ or $-1$ respectively.
Additionally, each orbital picks up $Q_{ii}=\pm1$ from the diagonal matrix $Q$ in orbital space [Eq.~\eqref{eq:q}].
Thus, $\ket{\psi} = \sum_{i} \psi_i(z) \ket{i}$ transforms to $\isopz\ket{\psi}=\sum_{i}\psi_i(-z) Q_{ii}\ket{i}=\sum_{i}p_iQ_{ii}\psi_i(z)\ket{i}$.
If all $p_iQ_{ii}$ are equal, $\ket{\psi}$ is an eigenstate of $\isopz$ with eigenvalue equal to $p_iQ_{ii}$.
For the state in Fig.~\ref{fig:ll}(b), the envelope functions for $\ket{\Gamma_6,+\frac{1}{2}}$, $\ket{\Gamma_8,+\frac{3}{2}}$, and $\ket{\Gamma_8,-\frac{1}{2}}$ ($Q_{ii}=+1$) are even, and the envelope functions for 
$\ket{\Gamma_6,-\frac{1}{2}}$, $\ket{\Gamma_8,+\frac{1}{2}}$, and $\ket{\Gamma_8,-\frac{3}{2}}$ ($Q_{ii}=-1$) are odd. The isoparity eigenvalue is thus $+1$. Analogously, it follows that Fig.~\ref{fig:ll}(c) shows a state with isoparity eigenvalue $-1$.
Thus, the mirror symmetry combined with the finite magnetic field gives rise to two distinct branches of LLs characterized by opposite isoparity quantum numbers.

In contrast to a finite magnetic field, a finite electrostatic field in $z$ direction leads to different densities in top and bottom surface and breaks the mirror symmetry.
In order to model the LL spectrum of Fig.~\ref{fig:ll}(d), we have assumed a potential drop $\Delta V=5\,\mathrm{meV}$ between the surfaces.
The wave functions are no longer even or odd under reflection in $z$, but bunch towards either the top or the bottom surface, see Figs.~\ref{fig:ll}(e) and (f). Thus, isoparity $\isopz$ is no longer a conserved quantum number.
As a result, the LL spectrum changes profoundly: Many crossings of LLs appear at finite magnetic fields.

\subsection{Effective model}

We may now exploit the property that the electrostatic field anticommutes with isoparity \cite{Isoparity,FundamentalSymmetries} to interpret the simultaneous presence of magnetic and electric fields in terms of competing isoparity-symmetric and antisymmetric terms.
Assuming that the fields are not too strong, we may conveniently model this competition in a perturbative model.
To the zero-field Hamiltonian $H_0$, which is isoparity-degenerate, we add two perturbative terms $H_\mathrm{S}$ and $H_\mathrm{A}$, symmetric and antisymmetric under isoparity,
\begin{equation}\label{eq:pert_model}
  H = H_0 + H_\mathrm{S} + H_\mathrm{A} \equiv H_0 + \alpha h_\mathrm{S} + \beta h_\mathrm{A}.
\end{equation}
By definition, the symmetric term commutes with the isoparity operator, $[H_\mathrm{S},\isopz]=0$, whereas the antisymmetric term anticommutes, $\{H_\mathrm{A},\isopz\}=0$.
We define the operators $h_\mathrm{S}$ and $h_\mathrm{A}$ to be dimensionless and of order $1$. The perturbative coefficients $\alpha$ and $\beta$ thus represent the respective energy scales of both terms. 

The out-of-plane magnetic field $\vec{B}=B\hat{z}$ transforms as an axial vector and is thus symmetric under $z\to-z$. As a symmetric perturbation, it thus enters into $H_\mathrm{S}$. It preserves isoparity, but lifts the degeneracy between the $\isopz=\pm 1$ eigenstates \cite{Isoparity}.
For the electric field, we have to decompose the electrostatic potential $V(z)$ into an even and an odd part, $V(z) = V_\mathrm{even}(z) + V_\mathrm{odd}(z)$ with $V_\mathrm{even}(-z) = V_\mathrm{even}(z)$ and $V_\mathrm{odd}(-z) = -V_\mathrm{odd}(z)$. The even part is related to total carrier density and commutes with $\isopz$.
The odd part, which induces a density imbalance between top and bottom, anticommutes with $\isopz$ \cite{Isoparity,FundamentalSymmetries} and constitutes the term $H_\mathrm{A}$.
Beyond the inversion symmetric approximation that we have adopted here for the Kane Hamiltonian, the bulk-inversion asymmetric terms would also contribute to the antisymmetric perturbation and enter into $H_\mathrm{A}$.

The spectrum of $H_0$ is two-fold degenerate: Its spectrum contains two copies of each energy eigenvalue $E_j$, where $j$ labels the band.
Since isoparity commutes with $H_0$, one can define a basis of isoparity eigenstates, denoted $\ket{\psi_{j,\pm}}$, for each two-dimensional degenerate eigenspace.
Here, isoparity can be interpreted as a type of pseudospin (not equal to proper spin or to total angular momentum).
The perturbative model of Eq.~\eqref{eq:pert_model} can be formulated in terms of the Pauli matrices $\sigma_x,\sigma_y,\sigma_z$, which thus encode the ``isoparity pseudospin'':
The isoparity-symmetric term can be parametrized as $\alpha_j\sigma_z$.
The isoparity-antisymmetric term can be written in terms of linear combinations of the Pauli matrices $\sigma_x$ and $\sigma_y$; we consider $\beta_j\sigma_x$ for simplicity and without loss of generality.
Thus, the perturbed Hamiltonian for the degenerate subspace (band $j$) is
\begin{equation}\label{eq:pert_model_minimal}
  H_j = E_j \idmat + \alpha_j\sigma_z + \beta_j\sigma_x.
\end{equation}
In the pseudospin picture, the (diagonal) term $\alpha_j\sigma_z$ acts as a Zeeman-type term, whereas the (off-diagonal) term $\beta_j\sigma_x$ resembles a Rashba-type effect.
From $H_j$, we find the energy eigenvalues (up to first order),
\begin{equation}\label{eq:pert_model_eigenvalues}
  E = E_j \pm \Delta E_j \qquad\text{with}\qquad \Delta E_j = \sqrt{\alpha_j^2 + \beta_j^2}.
\end{equation}
The eigenstates are given by
\begin{align}\label{eq:pert_model_eigenstates}
  \ket{\phi_{j,1}}
  &= \phantom{-}\cos \tfrac{1}{2}\theta\, \ket{\psi_{j,+}} + \sin \tfrac{1}{2}\theta\,\ket{\psi_{j,-}},\\
  \ket{\phi_{j,2}}
  &= -\sin \tfrac{1}{2}\theta\, \ket{\psi_{j,+}} + \cos \tfrac{1}{2}\theta\,\ket{\psi_{j,-}}\nonumber,
\end{align}
where $\theta$ is defined by $\tan \theta = \beta_j/\alpha_j$.
The isoparity expectation value of these states can straightforwardly be determined as
\begin{equation}\label{eq:pert_model_isoparity}
 \langle \isopz\rangle
 =\langle\sigma_z\rangle
 =\pm\cos\theta
 =\pm\frac{1}{\sqrt{1+\beta_j^2/\alpha_j^2}}
 =\pm\frac{\alpha_j}{\Delta E_j}.
\end{equation}
The latter expression provides a convenient method to calculate the isoparity expectation value indirectly from the energies, which we will employ later in order to estimate the isoparity expectation values from the experiment.

\begin{figure*}[t]
\includegraphics[width=178mm]{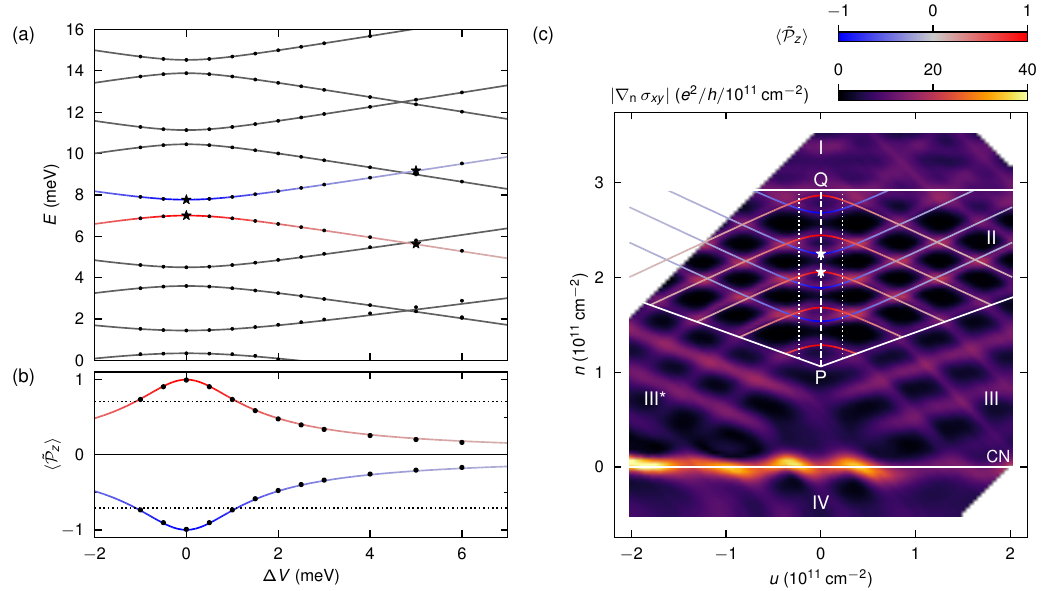}
\caption{\label{fig:model}%
Fit of the perturbation theory result to the numerical and experimental data.
(a) The Landau level energies calculated by \kdotp{} theory (dots) as function of the potential strength $\Delta V$ for a constant magnetic field $B=0.75\,\mathrm{T}$. Five pairs of Landau levels are shown.
The curves are fits of Eq.~\eqref{eq:pert_model_eigenvalues} to this data.
The stars correspond to the Landau levels highlighted in Fig.~\ref{fig:ll}.
For the highlighted pair of Landau levels, the color indicates the simulated isoparity value $\langle\isopz\rangle$ from Eq.~\eqref{eq:pert_model_isoparity}.
(b) Comparison of the simulated isoparity value (solid curves) with the expectation values $\langle\isopz\rangle$ of the eigenstates from \kdotp{} theory, for the highlighted pair of Landau levels.
The dotted lines indicate the crossover $\alpha=\beta$, i.e., where the isoparity symmetric and antisymmetric terms have the same energy scale.
(c) The experimental data has been plotted in terms of the coordinates $(u,n)$, corresponding to density imbalance and total density, respectively. This is achieved by appropriate scaling of the data in Fig.~\ref{fig:dualgate}(a) combined with a $45^\circ$ rotation.
We fit the functional dependence of Eq.~\eqref{eq:pert_model_density} to this data.
The colors of the curves indicate the isoparity expectation values $\langle\isopz\rangle$ extracted directly from the fitting parameters $\tilde{\alpha}$ and $s$.
The vertical dotted lines indicate the crossover $\alpha=\beta$.
Similar to (a), the stars highlight the Landau levels below and above the $\nu=12$ plateau.
}
\end{figure*}

We now verify the perturbation theory result against the Landau level energies calculated from $\mathbf{k}\cdot\mathbf{p}$ theory.
With the experiment in mind, we fix the magnetic field at $B=0.75\,\mathrm{T}$ and vary the potential strength $\Delta V$ (the potential difference between bottom and top surface).
The results of the \kdotp{} calculation are visualized as the data points (dots) in Fig.~\ref{fig:model}(a).
For each pair of Landau levels, we fit Eq.~\eqref{eq:pert_model_eigenvalues} to these data points, under the assumption that $\beta_j$ is proportional to $\Delta V$:
We determine $E_j$ and $\alpha_j$ from the limit $\Delta V\to 0$ ), where the energy eigenvalues are $E_j\pm\alpha_j$.
In the opposite limit ($\beta\gg\alpha$), the energies approach the linear dependence $E=E_j \pm \beta_j$. We thus determine the proportionality constant $s$ in $\beta_j=s\Delta V$ from the slope of the energy dependence at large $\Delta V$.
For the highlighted pair of Landau levels, the fit parameters are $E_j=7.38\,\mathrm{meV}$, $\alpha_j=0.38\,\mathrm{meV}$, and $s=0.35$.
We have plotted the fitted energy dependencies of Eq.~\eqref{eq:pert_model_eigenvalues} as curves in Fig.~\ref{fig:model}(a), and we observe good agreement between the perturbation theory and the $\mathbf{k}\cdot\mathbf{p}$ result. 

In Fig.~\ref{fig:model}(b), we compare the isoparity expectation values $\langle\isopz\rangle$ calculated from the $\mathbf{k}\cdot\mathbf{p}$ model (markers) with the perturbation theory prediction $\langle\isopz\rangle=\pm1/\sqrt{1+\beta^2/\alpha^2}$ (curves) for one pair of Landau levels.
The calculated isoparity dependence agrees well with the perturbation theory prediction, including the limiting behavior where  $\langle\isopz\rangle\to\pm 1$ in the limit $\Delta V=0\,\mathrm{meV}$ and $\langle\isopz\rangle\to0$ for large $\Delta V$.
The crossover between these limit occurs where the energy scales $\alpha_j$ and $\beta_j$ are of equal magnitude.
The isoparity values $\langle\isopz\rangle=\pm\frac{1}{2}\sqrt{2}$ 
where $|\beta_j/\alpha_j| = 1$ are visualized as the dotted lines in Fig.~\ref{fig:model}(b). In this case, the crossover occurs for $\Delta V=1.1\,\mathrm{meV}$.

We now return to the experiment.
We base the comparison of experiment to theory on the locations of the plateau transitions in the weighted-gradient data of region II in Fig.~\ref{fig:dualgate}(a).
To facilitate this comparison, we interpret the experimental data in terms of total density $n$ (along the line P--Q) and density imbalance $u$ (perpendicular to P--Q), see Fig.~\ref{fig:model}(c).
The coordinate transformation between $(\Vtg,\Vbg)$ and $(u, n)$ (see SI Appendix, section~\ref{sec:supp_data_analysis} for details) essentially rotates the data of Fig.~\ref{fig:dualgate}(a) by $45^\circ$, and places the $u$ and $n$ coordinate axes at the charge neutrality line and P--Q, respectively.  
We point out that the density imbalance coordinate $u$ merely parametrizes the system, and cannot generally be understood as a difference between densities of top and bottom surface state carriers: The latter quantities are ill-defined near the equal-density line where the eigenstates have probability density both at top and bottom simultaneously [cf. Fig.~\ref{fig:ll}(b) and (c)].
Instead, one may decompose the total carrier density $n$ as sum of densities $n_\mathrm{tg}+n_\mathrm{bg}$, where $n_\mathrm{tg}$ and $n_\mathrm{bg}$ are defined as the contributions due to top and bottom gate, respectively, but not as related to a specific carrier type. In this picture, $u= n_\mathrm{tg}-n_\mathrm{bg}$ is a valid interpretation as a density imbalance.

On a first (qualitative) glance at the transitions in region II in Fig.~\ref{fig:model}(c), we observe quadratic behavior for small and linear behavior for large $u$, consistent with the limits of small and large $\beta_j$ in Eq.~\eqref{eq:pert_model_eigenvalues}, respectively.
The quantitative comparison is based on the analogue of Eq.~\eqref{eq:pert_model_eigenvalues} in terms of carrier density $(u,n)$,
\begin{equation}\label{eq:pert_model_density}
 n_{j\pm}(u)
 = n_j \pm \sqrt{\tilde{\alpha}_j^2 + \tilde{\beta}^2}
 \qquad\text{with}\qquad \tilde{\beta}=su,
\end{equation}
where the offset $n_j$, half-splitting $\tilde{\alpha}_j$, and scale factor $s$ act as parameters, to be determined by fitting to the experimental data, in a similar manner as before:
We extract $n_j$ and $\tilde{\alpha}_j$ by identifying $n_{j\pm}(0)=n_j\pm\tilde{\alpha}_j$ from the intersects of the plateau transitions with the equal density line P--Q.
The scale factor $s$ can be extracted from the slopes of the transitions far from P--Q, where $n_{j\pm}(u)\approx n_j \pm s u$.
For the experimental data, we find that the single value $s=0.47$ fits well for all transitions.
For the highlighted set of transitions, we find $n_j=(2.15\pm0.05)\times10^{11}\,\mathrm{cm}^{-2}$ and $\tilde{\alpha}_j=(0.10\pm 0.02)\times10^{11}\,\mathrm{cm}^{-2}$. (The theoretical predictions are $12\,eB/h=2.18\times10^{11}\,\mathrm{cm}^{-2}$ and $\frac{1}{2}\,eB/h=0.091\times10^{11}\,\mathrm{cm}^{-2}$, respectively. The uncertainties of $n_j$ and $\alpha_j$ stem from uncertainties in positions of plateaus and in the estimated densities at P and Q.)
The resulting functions $n_{j\pm}(u)$ with the fitted parameters are plotted in Fig.~\ref{fig:model}(c), using the functional dependence of Eq.~\eqref{eq:pert_model_density} and the coordinate transformation as provided in SI Appendix, section~\ref{sec:supp_data_analysis}.  
If we would assume a constant density of states in the bulk gap of $0.104\times 10^{11}\,\mathrm{cm}^{-2}\,\mathrm{meV}^{-1}$ (due to the surface states), we find that $\tilde\alpha_j$ would correspond to an energy scale $\alpha= (1.0 \pm 0.2)\,\mathrm{meV}$. This value cannot be directly compared to $\alpha=(0.4\pm0.1)\,\mathrm{meV}$ from the $\mathbf{k}\cdot\mathbf{p}$ fit, because the density of states is far from constant.
In particular, the broadening of LLs due to thermal and disorder effects can severely affect the energy-to-density relation.

We determine $\langle \isopz\rangle$ as function of $u$ indirectly from the fit parameters $\tilde\alpha_j$ and $s$, using the identity $\langle \isopz\rangle=\pm1/\sqrt{1+ \tilde{\beta}^2/\tilde{\alpha}^2}$ (cf.\ Eq.~\eqref{eq:pert_model_isoparity}), which we validated in Fig.~\ref{fig:model}(b). 
The appropriate sign can be extracted from Figs.~\ref{fig:ll} and \ref{fig:model}(a), where $\langle \isopz\rangle<0$ for $n_{j+}(u)$ and $\langle \isopz\rangle>0$ for $n_{j-}(u)$.
The result is visualized in the red-white-blue color code in Fig.~\ref{fig:model}(c) and replicates the limiting behavior we have discussed before:
Near the equal-density line, we have isoparity eigenstates with $\langle \isopz\rangle=\pm1$. In the limit of large density imbalance, we obtain $\langle \isopz\rangle \to 0$, where the natural eigenstates have pure top or bottom surface-state character.

The crossover where the symmetric and antisymmetric contributions are equal in magnitude ($\tilde{\beta}=\tilde{\alpha}$) occurs where $u=u_\times\equiv\tilde\alpha/s$.
Averaging over all fitted transitions, we find $u_\times = (0.23\pm0.03)\times10^{11}\,\mathrm{cm}^{-2}$.
We have visualized this crossover with the dotted lines in Fig.~\ref{fig:model}(c).
Between the dotted lines, the effect of the magnetic field dominates and the eigenstates are (by approximation) isoparity eigenstates. Outside, the effect of the electric field ($V_\mathrm{odd}(z)$) dominates and the eigenstates are (by approximation) top and bottom states.
The gate voltage change required to reach $u=u_\times$ is $\Delta \Vtg=43\,\mathrm{mV}$ or $\Delta \Vbg=1.6\,\mathrm{V}$.

\subsection{Coexistence of massless and massive surface states}

In region III [the lower-right corner of Fig.~\ref{fig:dualgate}(a)], distinct sets of integer filling sequences are observed, with transitions of the total filling factor in steps of two. For example, $\nu = 3, 5, 7$ and $\nu = 4, 6, 8$ occur for increasing bottom gate voltage $\Vbg$ where the top gate voltage is kept constant at $\Vtg = -0.11\,\mathrm{V}$ and $-0.05\,\mathrm{V}$, respectively.
In this regime, the negative bottom gate voltage drives the Fermi level into the valence band at the bottom surface, which leads to Fermi level pinning and the formation of massive Volkov-Pankratov states \cite{VPS,MahlerVPS,Lixian}.
The steps of two in the vertical ($\Vbg$) direction are due to these $p$-type states at the bottom surface. Indeed, the $\mathbf{k}\cdot\mathbf{p}$ results of Fig.~\ref{fig:ll} indicate that the Landau levels for such states bunch up in pairs, consistent with Ref.~\cite{HoferFuchsTransport}.
At the same time,  the top of the device remains in the topological surface state regime, by virtue of efficient electrostatic screening \cite{MahlerVPS,Lixian}: The data in Fig.~\ref{fig:dualgate}(a) confirms that $\Vbg$ has almost no effect on the positions (values of $\Vtg$) of the plateau transitions. This confirms the coexistence of the ($p$-type) massive Volkov-Pankratov states and the massless (topological) surface states at the top surface, similar to the behavior reported in a previous study \cite{Lixian}.
Region III* [the upper-left region in Fig.~\ref{fig:dualgate}(a)] is fully analogous, with bottom and top exchanged. Here, $p$-type massive Volkov-Pankratov states at the top and topological surface states at the bottom side coexist.

\section{Discussion}

The experimentally observed pattern of Hall plateau transitions in the topological surface state regime (region II) can be accurately predicted with our simple perturbative model that describes the competition between isoparity symmetric and antisymmetric contributions.
The tunable amount of structural symmetry (isoparity) breaking in our dual-gated device is reflected in the observed behavior as we vary the carrier density by means of the top and bottom gates.
If the electrostatic potential is symmetric, the isoparity symmetric magnetic field dominates, leading to a splitting between two isoparity eigenstates.
These isoparity eigenstates are symmetric and antisymmetric combinations of the top and bottom surface states, so that the usual picture where the two topological surface states are independent is not valid in this regime.
They can only be treated as independent when the carrier density asymmetry is sufficiently strong, such that the isoparity-antisymmetric term dominates.
The crossover between the regimes is gradual and lies at an energy scale of $\sim 1\,\mathrm{meV}$, or equivalently, at a density imbalance of $\sim0.2\times10^{11}\,\mathrm{cm}^{-2}$. The gate voltage values corresponding to these values are strongly device dependent.
The quantitative estimate of the effects of the gates given a certain potential difference is an important piece of information in electrostatics simulations of such heterostructures, and aids the design of future devices.

Importantly, the default scenario in single-gate devices (usually top-gated) is the latter, symmetry-broken one: Upon tuning the top gate voltage, the electrostatic potential responds strongly at the top of the device, but only weakly at the bottom (see, e.g., Ref.~\cite{HoferEASelfcon}).
In 3D TI devices, the dominant effect of asymmetric gating is the density difference between the two surface states.
Both surface states are single spin states, because the combined crystalline and structural symmetry `entangles' the top-or-bottom degree of freedom with the spin direction.
In contrast, in two-dimensional systems asymmetric gating induces the Rashba effect, which manifests as a splitting between a pair of spin states.
For a topological surface state in a 3D TI, the lack of a second spin state prohibits the Rashba mechanism from occurring.
The implications of an applied potential difference between top and bottom surface are consequently more subtle: The amount of structural inversion symmetry breaking affects the mixing of the eigenstates, and in general, the surface states cannot be viewed as independent.

\begin{acknowledgments}
We thank C.~Ames for growing the material and J.~Kleinlein for supervising the fabrication of the device.
We acknowledge financial support from the Deutsche Forschungsgemeinschaft (DFG, German Research Foundation) in the project SFB 1170 (Project ID 258499086) and in the W\"urzburg-Dresden Cluster of Excellence on Complexity, Topology and Dynamics in Quantum Matter \textit{ctd.qmat} (EXC 2147, Project ID 390858490).
\end{acknowledgments}

\section*{Author contributions}
H.B. and L.W.M. designed the experiment.
V.L.M. fabricated the device.
S.S.Y. and V.L.M. performed the transport experiments.
S.S.Y., V.L.M., and W.B. analyzed the experimental data.
W.B., M.H., M.S., and T.K. performed the numerical analysis.
W.B. and E.M.H. developed the theoretical model.
W.B., T.K., E.M.H., H.B., and L.W.M. supervised the project.
W.B. led the writing of the manuscript, with input provided by all authors.

\section*{Data availability}
The data that support the findings of this article are publicly available \cite{dataset_v2}.



%


\clearpage

\setcounter{figure}{0}
\setcounter{section}{0}
\renewcommand{\thefigure}{S\arabic{figure}}
\renewcommand{\theHfigure}{S\arabic{figure}}  
\renewcommand{\thesection}{\Alph{section}}
\renewcommand{\figurename}{\textsc{Supplementary Figure}}


\onecolumngrid
\noindent
{\centering\large\bf Supplementary Information for\\``Symmetry induced spin splitting of topological surface states at equal carrier density''\\}
\bigskip

\section{Device fabrication and characterization}
\label{sec:supp_device}

We use molecular beam epitaxy (MBE) to grow the material on top of an approximately $3\,\mathrm{mm}\times 2\,\mathrm{mm}$ sized Si-doped GaAs substrate. The substrate is covered with a thick CdTe buffer layer.
The sample stack is formed by the $80\,\mathrm{nm}$ thick Hg$_{0.3}$Cd$_{0.7}$Te bottom barrier, the $70\,\mathrm{nm}$ HgTe layer, and the $80\,\mathrm{nm}$ thick Hg$_{0.3}$Cd$_{0.7}$Te top barrier.
The top-gate stack consists of an insulating dielectric layer of $15\,\mathrm{nm}$ thick HfO$_\mathrm{x}$ and a $10\,\mathrm{nm}/100\,\mathrm{nm}$ Ti/Au gate electrode.
The GaAs substrate acts as the bottom gate.

\begin{figure}
 \centering
 \includegraphics[width=178mm]{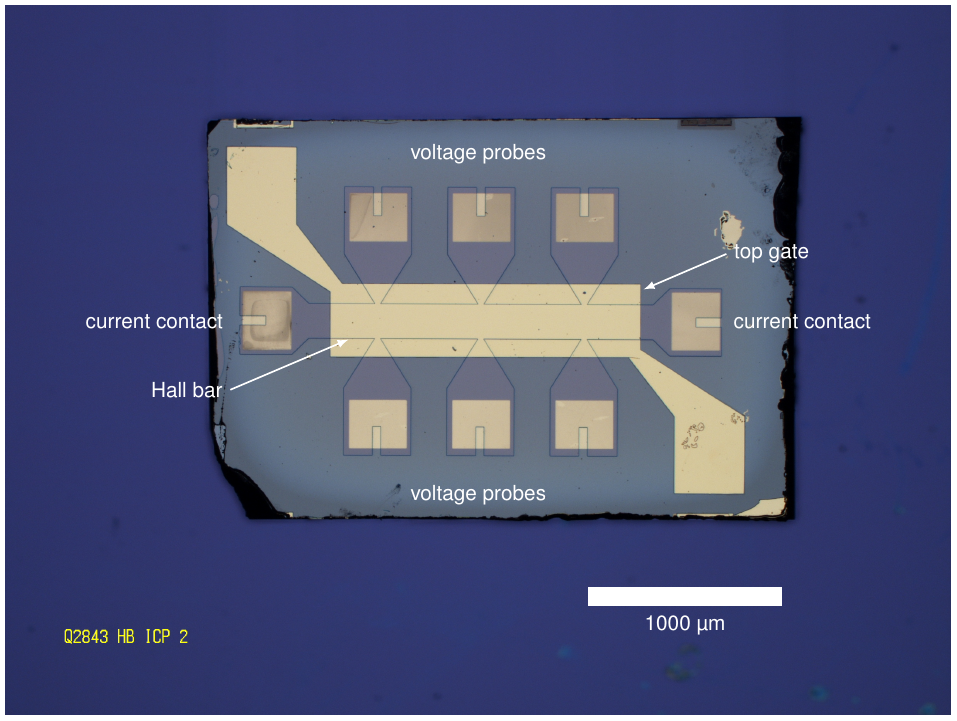}
 \caption{\label{figsupp:deviceimage}
 Optical image of the Hall bar device, top view.
 The relevant components in the device are indicated in the image.
 The distance between the voltage probes is $\SI{540}{\micro\meter}$ and the width of the Hall bar is $\SI{180}{\micro\meter}$.
 }
\end{figure}

We have etched an eight terminal Hall bar structure (see Fig.~\ref{figsupp:deviceimage}) with the inductively coupling plasma (ICP) etching technique using Ar. The distance between two neighboring voltage contacts is $\SI{540}{\micro\meter}$ and the width of the Hall bar is $\SI{180}{\micro\meter}$, giving an aspect ratio (length divided by width) of $3$.

The carrier mobility is $\mu\approx 600\times10^{3}\,\mathrm{cm}^2\,\mathrm{V}^{-1}\,\mathrm{s}^{-1}$ in the $n$-type regime (total carrier density $n\gtrsim 1\times 10^{11}\,\mathrm{cm}^{-2}$).
In the $p$-type regime, the mobility is of the order of $\sim100\times10^{3}\,\mathrm{cm}^2\,\mathrm{V}^{-1}\,\mathrm{s}^{-1}$ for $n= -5\times 10^{11}\,\mathrm{cm}^{-2}$. These mobility values have been determined from single-carrier Drude fits and thus contain contributions from all carriers at the given densities.

\section{Magnetotransport experiments}
\label{sec:supp_transport}

We have performed the magnetotransport measurements in the usual Hall bar configuration, measuring longitudinal $V_{xx}$ and transversal voltage $V_{xy}$ across two neighboring contact pairs as indicated in Fig.~\ref{fig:sampleraw} of the main text. 
We measure these voltages using a standard low-frequency AC lock-in technique at  $37\,\mathrm{Hz}$. The current $I$ through the device is determined from the voltage drop across a calibrated resistor in series.
The experiments have been performed in a dilution refrigerator at its nominal base temperature of $150\,\mathrm{mK}$.

Each measurement run has been performed at a given constant perpendicular magnetic field.
For values of the bottom gate voltage $\Vbg$ from $-10\,\mathrm{V}$ to $10\,\mathrm{V}$, increased in steps of $0.2\,\mathrm{V}$, we have swept the top gate voltage $\Vtg$ from $-0.8\,\mathrm{V}$ to $0.2\,\mathrm{V}$ in steps of $5\,\mathrm{mV}$.

\section{Data analysis}
\label{sec:supp_data_analysis}

\begin{figure}
 \centering
 \includegraphics[width=84mm]{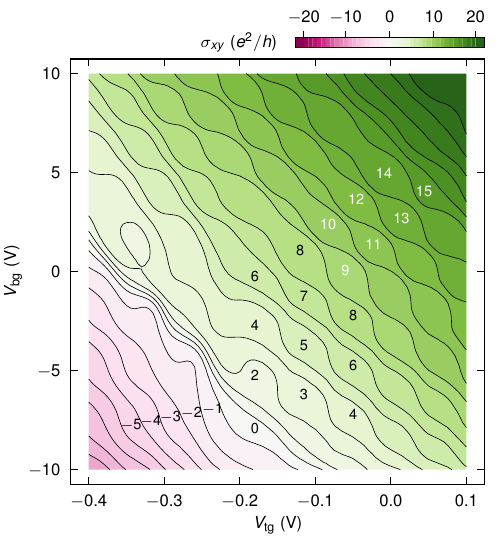}
 \caption{\label{figsupp:sigmaxy}
 Hall conductivity extracted from the measured resistances.
 The Hall conductance $\sigma_{xy}$ calculated from the longitudinal ($R_{xx}$) and transversal ($R_{xy}$) resistance. The thin contours are at half-integer multiples of $e^2/h$. The integer values indicate the multiples $\nu$ in $\sigma_{xy}=\nu e^2/h$. Values have been rounded to the nearest integer.
 }
\end{figure}

The Hall conductivity $\sigma_{xy}$ is determined from the longitudinal and transversal resistances as
$\sigma_{xy} = R_{xy} / [(R_{xx}/r)^2+R_{xy}^2]$.
where $r=3$ is the geometric aspect ratio.
We have plotted $\sigma_{xy}(\Vtg,\Vbg)$ for top gate voltages $\Vtg$ in the range $-0.4\,\mathrm{V}$ to $0.1\,\mathrm{V}$ (where the data outside this range is not relevant for the discussion) and the bottom gate voltage $\Vbg$ in the range $-10\,\mathrm{V}$ to $10\,\mathrm{V}$, see Fig.~\ref{figsupp:sigmaxy}.
The gate voltage ranges for $\Vtg$ and $\Vbg$ span approximately the same densities: The gate actions $A_\mathrm{tg}$ and $A_\mathrm{bg}$, defined as the change in density per unit of gate voltage for both gates, differ by a factor of $A_\mathrm{tg}/A_\mathrm{bg}\approx40$.

\begin{figure}
 \centering
 \includegraphics[width=178mm]{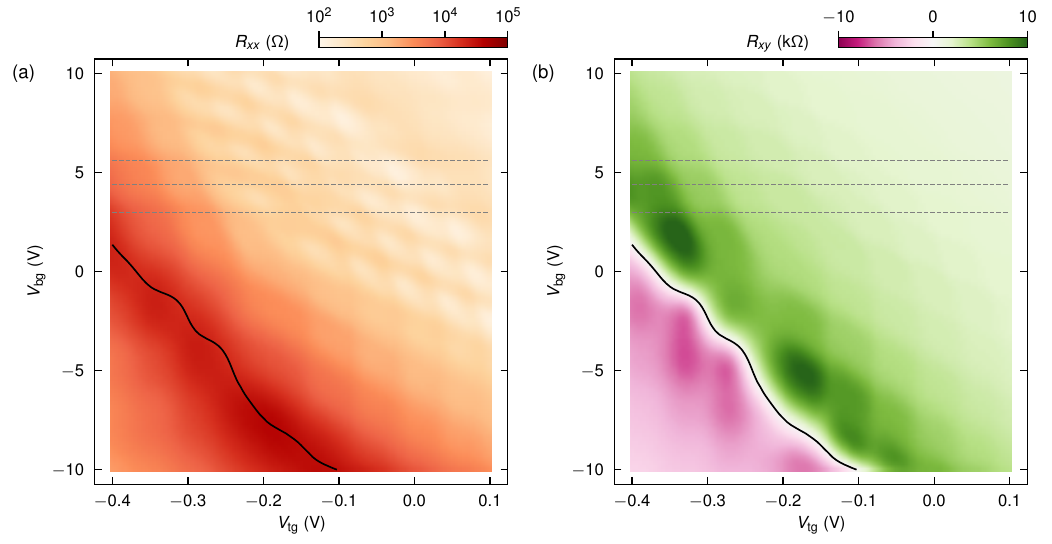}
 \caption{\label{figsupp:RxxRxy}
 Measured longitudinal and transversal resistance.
 (a) The measured longitudinal resistance $R_{xx}$ in a magnetic field of $B=0.75\,\mathrm{T}$. Note the logarithmic color scale.
 (b) The measured transversal resistance $R_{xy}$.
 The thick line indicates the boundary between negative and positive values of $R_{xy}$.
 The dashed lines indicate the bottom gate voltages $V_\mathrm{bg}=3.0$, $4.4$, and $5.6\,\mathrm{V}$ for which $R_{xx}$ and $R_{xy}$ are shown in Figs.~\ref{fig:sampleraw}(b) and (c) of the main text.
 }
\end{figure}

\begin{figure}
 \centering
 \includegraphics[width=178mm]{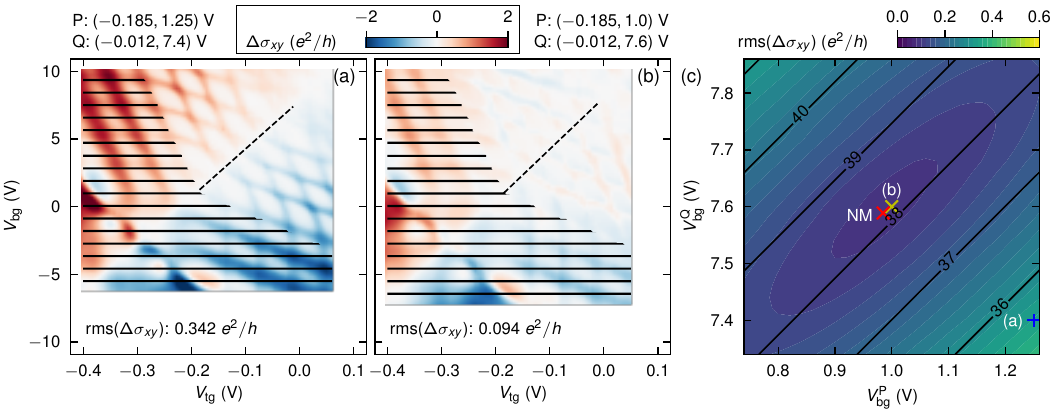}
 \caption{\label{figsupp:symmdiff}
 Optimization of the locations of P and Q.
 In the optimization procedure, the top gate voltages $\Vtg$ of P and Q are fixed and optimization is done for the bottom gate voltages $\Vbg$ of P and Q.
 (a) Symmetrization error $\Delta\sigma_{xy}$ for the initial estimate of P and Q. The data are scaled such that the line P--Q is at an angle of $45^\circ$ with the coordinate axes, see Appendix C for details. We indicate the root mean square (rms) value of $\Delta\sigma_{xy}$, excluding the hatched area. 
 (b) Symmetrization error $\Delta\sigma_{xy}$ for the optimized values of P and Q.
 (c) The rms difference of the original and reflected $\sigma_{xy}$ [cf.~(a) and (b)] as function of the $\Vbg$ values of P and Q, visualized by the shading.
 The optimization requires finding the minimum of this function.
 The blue plus marker corresponds indicates the initial estimate (a). The yellow cross is the optimized value (b), equal to the minimum value of $\mathrm{rms}(\Delta\sigma_{xy})$ on the grid of values $(\Vbg^\mathrm{P}, \Vbg^\mathrm{Q})$ for which this function was evaluated. The red cross is the minimum found by the Nelder-Mead (NM) method.
 The numbered black lines indicate the ratio of gate actions, $A_\mathrm{tg}/A_\mathrm{bg}$.
 }
\end{figure}

For the plot of the weighted gradient $\nabla_n\sigma_{xy}(\Vtg,\Vbg)$ [see Eq.~\eqref{eq:weighted_gradient} and Fig.~\ref{fig:dualgate}(a) of the main text], we have scaled the $\Vtg$ and $\Vbg$ axes based on $A_\mathrm{tg}$ and $A_\mathrm{bg}$, such that the equal density line through P and Q is at an angle of exactly $45^\circ$ with the axes.  
The gate voltage coordinates of P and Q are initially estimated by visual inspection and are subsequently optimized by demanding that $\sigma_{xy}(\Vtg,\Vbg)$ is as symmetric as possible by reflection over the line P--Q.
The initial estimates are $(\Vtg,\Vbg) = (-0.185, 1.25)\,\mathrm{V}$ for P and $(-0.012, 7.4)\,\mathrm{V}$ for Q.
Given P and Q, we stretch the $\Vbg$ axis such that P--Q is at $45^\circ$ and perform a perpendicular reflection of $\sigma_{xy}(\Vtg,\Vbg)$, the result of which is written as $R_\mathrm{PQ}(\sigma_{xy})$.
We optimize by minimizing the mean squared error $\mathop{\mathrm{rms}}(\Delta\sigma_{xy})$, defined by
\begin{equation}\label{eq:mean_squared_error}
  \mathop{\mathrm{rms}}(\Delta\sigma_{xy})^2=\frac{1}{A}\int |R_\mathrm{PQ}(\sigma_{xy}) - \sigma_{xy} |^2 \mathrm{d}\Vtg\mathrm{d}\Vbg,
\end{equation}
by adjusting the $\Vbg$ coordinates of P and Q in small steps, while keeping their $\Vtg$ coordinates fixed.
(In Eq.~\eqref{eq:mean_squared_error}, $A$ is the area of the domain being integrated over.)
The result is $(\Vtg,\Vbg) = (-0.185, 1.0)\,\mathrm{V}$ for P and $(-0.012, 7.6)\,\mathrm{V}$ for Q; see Fig.~\ref{figsupp:symmdiff} for an illustration.
It follows that the ratio of gate actions is $A_\mathrm{tg}/A_\mathrm{bg}=(\Vbg^\mathrm{P} - \Vbg^\mathrm{Q})/(\Vtg^\mathrm{P} - \Vtg^\mathrm{Q})\approx 38$.
Figs.~\ref{fig:dualgate}(a) and \ref{fig:model}(c) of the main text, as well as Figs.~\ref{figsupp:RxxRxy} and \ref{figsupp:sigmaxy} of the Supporting Information use these optimized values for P and Q, with the appropriate scaling of the axes such that P--Q runs at an exactly $45^\circ$ angle. 

In order to interpret the gate voltage coordinates in terms of densities, we first determine the total density $n$ along the equal-density line P--Q.
To this aim, we first define an auxiliary coordinate system $(X,Y)$, such that $(X,Y) = (0,0)$ and $(0,1)$ correspond to P and Q.
The transformation that links $(X,Y)$ to the top and bottom gate voltages $(\Vtg,\Vbg)$ is
\begin{align}\label{eq:coordtrans_y}
  \Vtg &= \Vtg^\mathrm{P} + \left(\Vtg^\mathrm{Q} - \Vtg^\mathrm{P}\right)(Y+X)\nonumber\\
  \Vbg &= \Vbg^\mathrm{P} + \left(\Vbg^\mathrm{Q} - \Vbg^\mathrm{P}\right)(Y-X),
\end{align}
in terms of the gate voltage coordinates of P and Q.
We find that the charge neutrality point, i.e., where the line P--Q intersects the boundary between regions III and IV [cf. Fig.~\ref{fig:dualgate}(a) of the main text], is located at $(0,Y_\mathrm{CNP})$ with $Y_\mathrm{CNP}=-0.57\pm0.02$.  

\begin{figure}
 \centering
 \includegraphics[width=84mm]{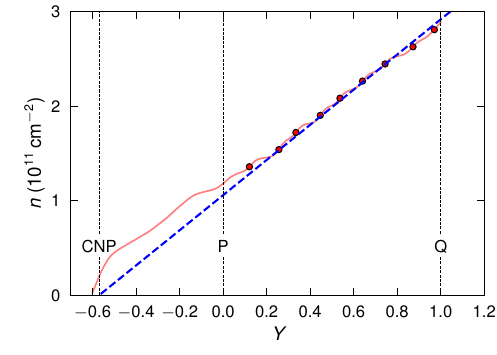}
 \caption{\label{figsupp:densityfit}
 Extraction of carrier density by fitting Hall conductivity.
 From $\sigma_{xy}$ along the line P--Q we find the intersection points $Y_j$ for which $\sigma_{xy}=(j+\frac{1}{2})e^2/h$ for integer $j$ (red dots).
 We fit $n=a(Y-Y_\mathrm{CNP})$ to the values $(Y_j, n_j)$ ($n_j = (j+\frac{1}{2})eB/h$) by least squares and find the slope $a=1.86\times10^{11}\,\mathrm{cm}^{-2}$.
 The red curve is the scaled Hall conductance $\sigma_{xy}\times eB/h$ from the measurement, shown as guide to the eye.
 }
\end{figure}

The relation between the $Y$ coordinate and total carrier density $n$ is inferred from fitting to the Hall conductivity $\sigma_{xy}$ along the line P--Q [cf.\ Fig.~\ref{fig:dualgate}(b) of the main text], as follows.  
We numerically solve $\sigma_{xy}(Y_j)=j+\frac{1}{2}$ for integers $j=7,8,\ldots 15$, which yield the positions $Y_j$ of the transitions between Hall plateaus.
The corresponding density values are $n_j=(j+\frac{1}{2})eB/h$, with $eB/h=0.181\times10^{11}\,\mathrm{cm}^{-2}$ for $B=0.75\,\mathrm{T}$.
We then fit the function $a(Y-Y_\mathrm{CNP})$ by least squares (with fit parameter $a$) to the data points $(Y_j, n_j)$, see Fig.~\ref{figsupp:densityfit}.
The fitted value is $a=(1.86\pm0.03)\times 10^{11}\,\mathrm{cm}^{-2}$ and P and Q correspond to a total density $\nP= (1.06\pm0.02)\times10^{11}\,\mathrm{cm}^{-2}$ and $\nQ = (2.92\pm0.02)\times10^{11}\,\mathrm{cm}^{-2}$, respectively.
(The uncertainty in $a$ is from the fit and from $Y_\mathrm{CNP}$. The uncertainties in $\nP$ and $\nQ$ are from the uncertainties in $a$ and in $Y_\mathrm{CNP}$.)

This information suffices to establish the coordinate transformation between $(\Vtg,\Vbg)$ on the one hand and density coordinates $(u,n)$ on the other hand.
We define the virtual carrier densities $n_\mathrm{tg}$ and $n_\mathrm{bg}$ as being linear in the gate voltages $\Vtg$ and $\Vbg$, respectively,
\begin{align}\label{eq:coordtrans_ntg_nbg}
  n_\mathrm{tg} &= \frac{1}{2}\left(\nP + \frac{\nQ - \nP}{\Vtg^\mathrm{Q} - \Vtg^\mathrm{P}}(\Vtg-\Vtg^\mathrm{P})\right)\nonumber\\
  n_\mathrm{bg} &= \frac{1}{2}\left(\nP + \frac{\nQ - \nP}{\Vbg^\mathrm{Q} - \Vbg^\mathrm{P}}(\Vbg-\Vbg^\mathrm{P})\right).
\end{align}
The two ratios of the form $(\nQ - \nP)/(V^\mathrm{Q}-V^\mathrm{P})$ are the estimated gate actions $A_\mathrm{tg}$ and $A_\mathrm{bg}$ respectively.
The total carrier density $n$ and the density imbalance $u$, with
\begin{equation}\label{eq:coordtrans_n}
  n = n_\mathrm{tg} + n_\mathrm{bg},\qquad
  u = n_\mathrm{tg} - n_\mathrm{bg}
\end{equation}
form the rotated coordinate system $(u, n)$ used for the comparison between model and experiment [see Fig.~\ref{fig:model}(c) of the main text in particular].  

\section{Numerical analysis}
\label{sec:supp_numerics}

For the $\mathbf{k}\cdot\mathbf{p}$ calculations with the band structure package \texttt{kdotpy} we have used the established material parameters for (Hg,Cd)Te and HgTe \cite{Novik,kdotpy}, with the following changes \cite{Brunescreening,ZieglerQHE,HoferFuchsTransport}: The Kane parameter $P$ is given by $E_\mathrm{P}=2m_0P^2/\hbar^2=21.0\,\mathrm{eV}$ and the valence band offset between HgTe and CdTe has been set to $-620\,\mathrm{meV}$.
The Landau level index cutoff is $n_\mathrm{LL,max}=30$. We have included non-axial and bulk inversion asymmetric contributions, with the standard coefficient values \cite{kdotpy}.
The broadening applied to the density of states has a Gaussian shape with a standard deviation of $0.75\,\mathrm{meV} \times \sqrt{B / B_0}$ where $B_0 \equiv 1\,\mathrm{T}$.

We apply a piecewise linear electrostatic potential $V(z)$, consistent with a constant displacement field $\vec{D}=\epsilon(z)\vec{E}(z)$, where $\vec{E}(z) = E(z)\hat{z}$ is the electric field and $\epsilon(z)$ the relative dielectric constant as function of $z$.
The nominal strength of the potential is given by $\Delta V=V(z_\mathrm{top}) - V(z_\mathrm{bottom})$, i.e., the potential difference between the top and bottom interface of the HgTe layer.
For Fig.~\ref{fig:model}(a) of the main text, we sweep $\Delta V$ from $-1\,\mathrm{meV}$ to $3\,\mathrm{meV}$ in steps of $0.5\,\mathrm{meV}$ and up to $6\,\mathrm{meV}$ in steps of $1\,\mathrm{meV}$ and extract the LL energies and eigenstate properties at the constant field of $B=0.75\,\mathrm{T}$.  
We have not applied the self-consistent Schr\"odinger-Poisson solver for this analysis, because the precise potential profile does not qualitatively affect the conclusions. If we would have used it, we would observe some small, but insignificant quantitative changes, but at the expense of much larger calculation times.

\begin{figure}
 \centering
 \includegraphics[width=178mm]{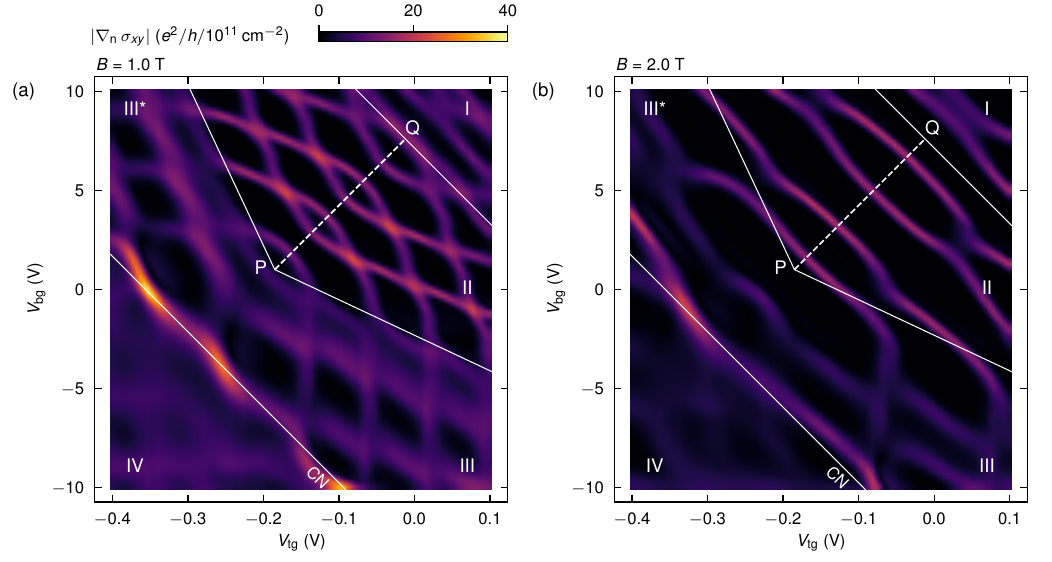}
 \caption{\label{figsupp:other_magn_fields}
 Data for other magnetic fields.
 (a) $B=1.0\,\mathrm{T}$.
 (b) $B=2.0\,\mathrm{T}$.
 The white lines that separate the transport regimes are placed at the same positions as for $B=0.75\,\mathrm{T}$, see Fig.~\ref{fig:dualgate}(a) of the main text.
 }
\end{figure}

\end{document}